\documentclass{svproc}
\usepackage{url}
\usepackage{graphicx}

\newcommand{\iso}[2]{{}^{#2}\mbox{#1}}

\begin{document}
\mainmatter              
\title{The s-process nucleosynthesis in low mass
stars: \\impact of the uncertainties in the nuclear physics
determined by Monte Carlo variations}
\titlerunning{The s-process in low mass stars}  
%
\author{Gabriele Cescutti \inst{1} \and
  Raphael Hirschi \inst{2,3} \and
 Nobuya Nishimura \inst{4} \and
 Thomas Rauscher \inst{ 5,6} \and
Jacqueline den Hartogh \inst{ 2,7} \and
 Alex St. J. Murphy \inst{8} \and
Sergio Cristallo \inst{9}}
\authorrunning{Gabriele Cescutti et al.} 

\institute{INAF, Osservatorio Astronomico di Trieste, Italy \\
\email{gabriele.cescutti@inaf.it}
\and
Astrophysics group, Faculty of Natural Sciences, Keele University, UK \\
\and
Kavli IPMU (WPI), University of Tokyo, Japan
\and
Yukawa Institute for Theoretical Physics, Kyoto University, Japan\\
\and
Department of Physics, University of Basel, Switzerland\\
\and
Centre for Astrophysics Research, University of Hertfordshire, UK\\
\and
Konkoly Observatory, Budapest, Hungary\\
\and SUPA, School of Physics and Astronomy, University of Edinburgh, UK\\
\and Osservatorio Astronomico d'Abruzzo,Teramo, Italy\\
\and INFN - Sezione di Perugia,  Italy
}

\maketitle              

\begin{abstract}
  We investigated the impact of uncertainties in neutron-capture and
  weak reactions (on heavy elements) on the s$-$process
  nucleosynthesis in low-mass stars using a Monte-Carlo based
  approach.  We performed extensive nuclear reaction network
  calculations that include newly evaluated temperature-dependent
  upper and lower limits for the individual reaction rates. Our
  sophisticated approach is able to evaluate the reactions that impact
more significantly the final abundances. We found
  that $\beta$-decay rate uncertainties affect typically nuclides
  near s-process branchings, whereas most of the uncertainty in the
  final abundances is caused by uncertainties in neutron capture
  rates, either directly producing or destroying the nuclide of
  interest. Combined total nuclear uncertainties due to reactions on
  heavy elements are approximately  50\%.

\end{abstract}
\section{Introduction}
\begin{figure}[ht!]
\begin{center}
\includegraphics[width=0.8\textwidth]{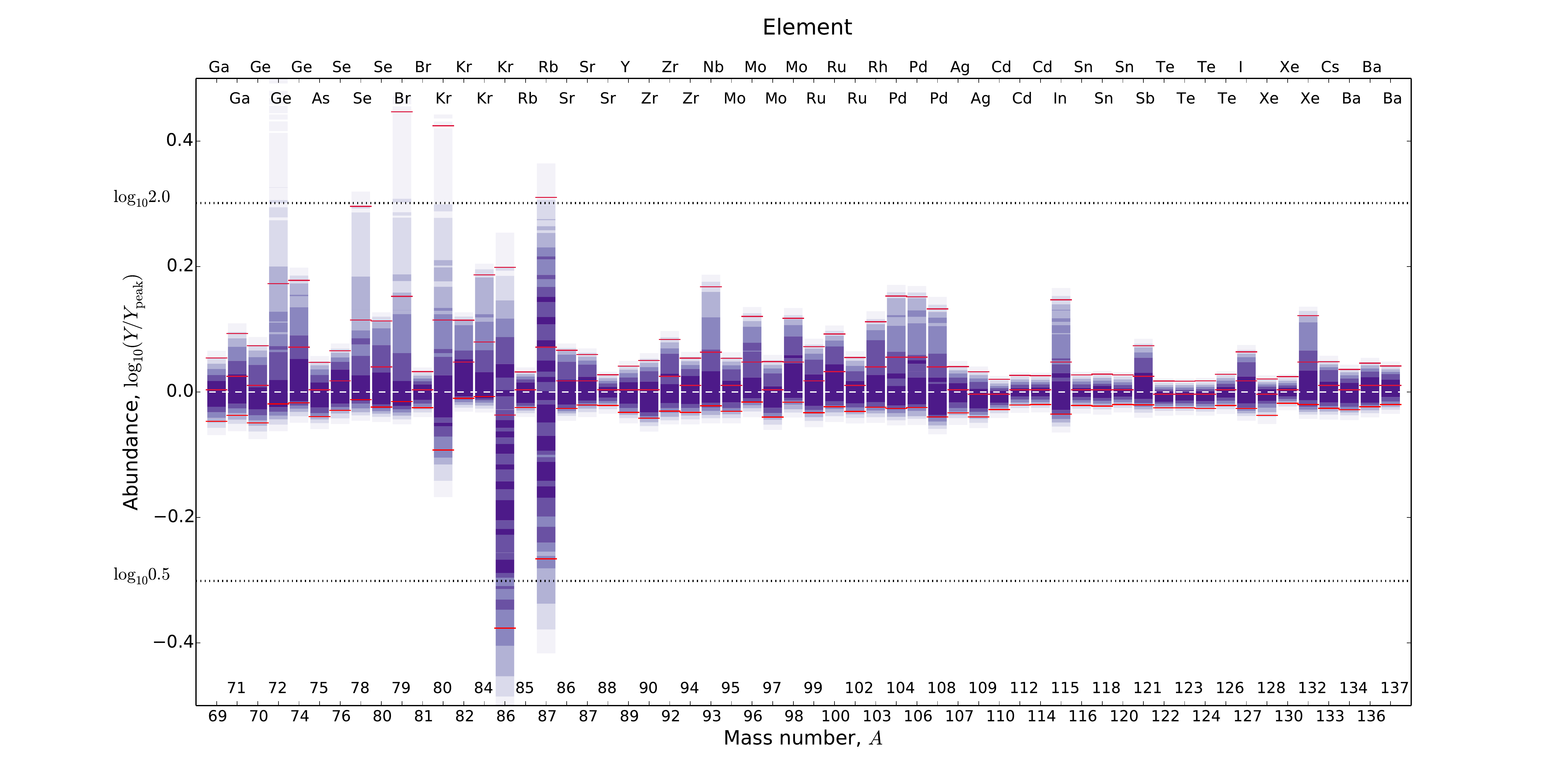}
\includegraphics[width=0.8\textwidth]{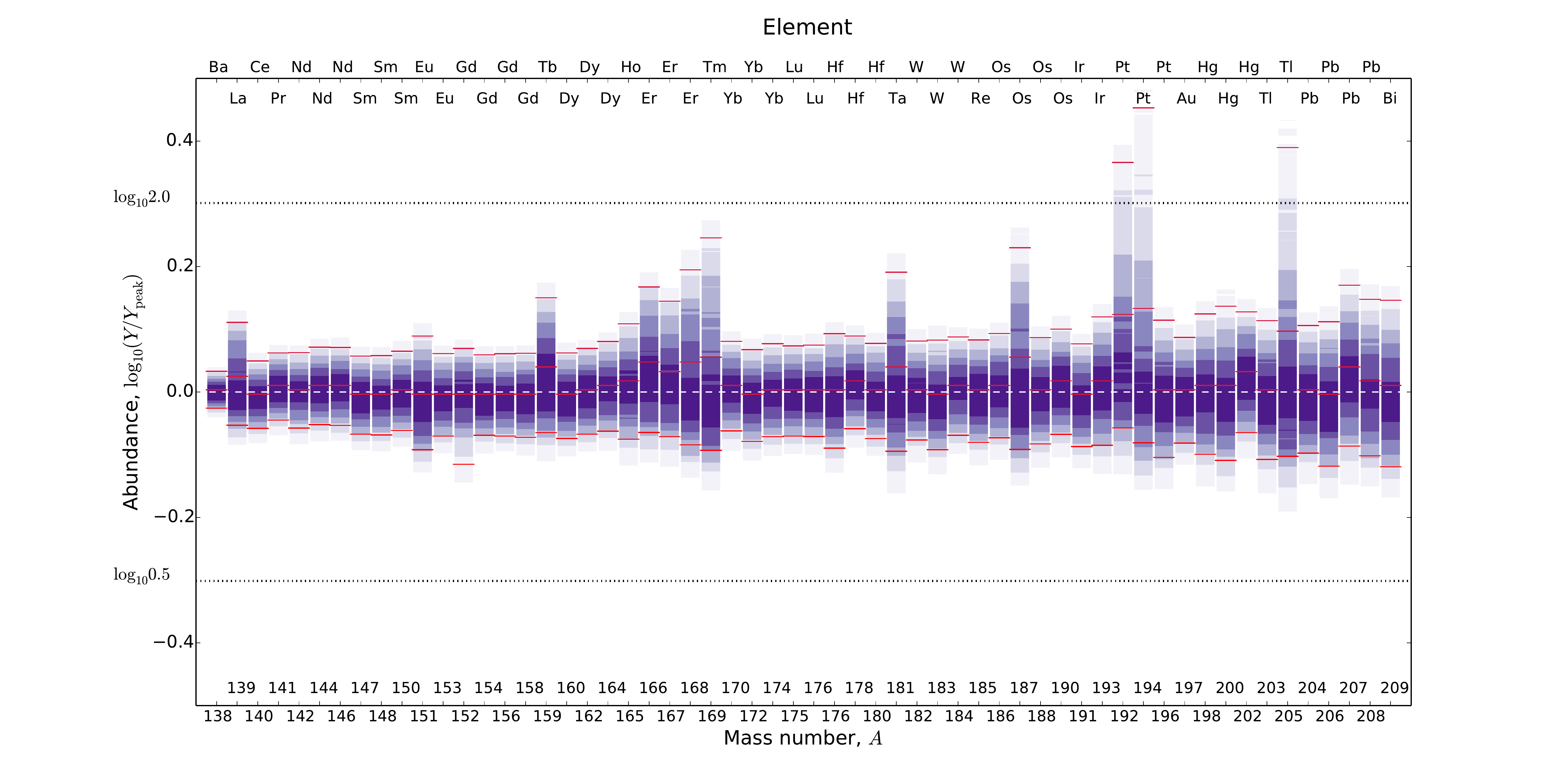}

\includegraphics[width=5cm]{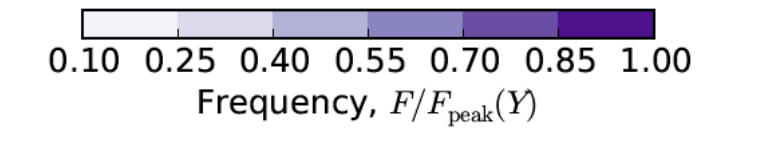}
\vspace{-.3cm}

\caption{Total production uncertainties in the s-process
  abundances at the end of a thermodynamic trajectory 
  approximating a  $^{13}$C pocket in a 3M$_{\odot}$ stars of solar
  metallicity. The color shading denotes the probabilistic frequency and the
 90\% probability intervals up and down are marked for each nuclide
 with the red lines. 
The final abundances are normalised by the final abundance at the peak of the distribution.
Horizontal dotted lines indicate a factor of two
uncertainties.}\label{L1}
\vspace{-.85cm}

\end{center}
\end{figure}
The s-process nucleosynthesis is a source of heavy elements beyond
iron in the universe, taking place in stellar burning
environments. There are two astronomical conditions and corresponding
classes of the s-process. The s-process occurs (i) during the AGB phase
of low mass stars producing heavy nuclei up to Pb and Bi, called
the main s-process; (ii) in He-core and C-shell burning phases of massive
stars representing the lighter components (up to $A \approx 90$),
categorised as the weak s-process.  Here, we investigate the main
s-process production in low-mass AGB stars. There are several well
known uncertainties concerning this production.  In this work, we
explore the nuclear reaction side, in particular the uncertainties in
neutron captures and beta decays on intermediate and heavy
isotopes. Our approach is to vary simultaneously all reaction rates in
a Monte Carlo (MC) framework rather than one reaction at a time. We
followed the same procedure as presented in detail in
\cite{Rauscher16}.  Furthermore, we use temperature-dependent
uncertainties based both on experimental and theoretical studies as we
have already done for several other processes: the s-process in
massive star, $\gamma$-process in core collapse SNe and
$\gamma$-process in supernovae type Ia
(\cite{Nishimura17,Rauscher16,Nishimura18}.)  On the astrophysical
side, the evolution of low-mass stars is complex, especially
during the TP-AGB phase (\cite{Cristallo15}). It is thus not feasible to repeat such
simulations 10000 times as required by the MC procedure to complete a
sensitivity study. We thus have to approximate the thermodynamic
conditions inside the star with a trajectory following the key phase
that we are studying (for details see \cite{Cescutti18}).

\section{Results}

As can be seen in Fig.\,\ref{L1}, the overall uncertainties at the end
of the trajectory approximating a $\iso{C}{13}$ pocket in a 3
M$_{\odot}$ star of solar metallicity are generally small. Indeed,
most of them are smaller than 50\%. This is not too surprising since
the relevant temperature range ($\sim$8\,keV) is accessible to
experimental measurements so many of the relevant rates. There are
nevertheless several nuclides, for which uncertainties are larger than
a factor of two. These are generally nuclides around branching points
such as $\iso{Kr}{86}$. We also notice a propagation effect for
nuclides more massive than $\iso{Ba}{138}$. This is due to the
combined effect of uncertainties in neutron capture rates above
$\iso{Ba}{138}$.  In most cases, rates dominating the nuclear
uncertainties are the neutron captures either directly producing or
destroying the nuclide in question (for the full list see
\cite{Cescutti18}). There are, however, three neutron-capture rates
that play a significant role in the uncertainty for many nuclides
during the $^{13}$C-pocket conditions. These are the neutron capture
rates on $\iso{Fe}{56}$, $\iso{Ni}{64}$, and $\iso{Ba}{138}$.  For a
detailed analysis of the how the importance of these key rates are
determined by examining the correlation between a change in a reaction
rate and the change of an abundance, we refer the reader to
\cite{Cescutti18}.

\paragraph{Acknowledgments}
GC acknowledges financial support from the European Union Horizon 2020
research and innovation programme under the Marie Skłodowska-Curie
grant agreement no. 664931.  This work has been partially supported by
the European Research Council (EU-FP7-ERC-2012-St Grant 306901), the
EU COST Action CA16117 (ChETEC), the UK STFC (ST/M000958/1), and MEXT
Japan (Priority Issue on Post-K computer: Elucidation of the
Fundamental Laws and Evolution of the Universe).

%
%

\end{document}